\documentclass[aps,prb,superscriptaddress,twocolumn]{revtex4}

\usepackage{graphicx}
\usepackage{color}

\begin{document}


\title{Enhanced spin-flip scattering at the surface of copper in lateral spin valves}


\author{Mikhail Erekhinsky}
\email[]{merekhinsky@physics.ucsd.edu}
\affiliation{Department of Physics, University of California-San Diego, La Jolla California 92093-0319, USA}

\author{Amos Sharoni}
\affiliation{Department of Physics, University of California-San Diego, La Jolla California 92093-0319, USA}

\author{F\`elix Casanova}
\affiliation{Department of Physics, University of California-San Diego, La Jolla California 92093-0319, USA}
\affiliation{Nanodevices Laboratory, CIC nanoGUNE Consolider, 20018 Donostia-San Sebastian, Basque Country}

\author{Ivan K. Schuller}
\affiliation{Department of Physics, University of California-San Diego, La Jolla California 92093-0319, USA}

\date{\today}

\begin{abstract}
We performed non-local electrical measurements of a series of Py/Cu lateral spin valve devices with different Cu thicknesses. We show that both the spin diffusion length of Cu and the apparent spin polarization of Py increase with Cu thickness. By fitting the results to a modified spin-diffusion model, we show that the spin diffusion length of Cu is dominated by spin-flip scattering at the surface. In addition, the dependence of spin polarization of Py on Cu thickness is due to a strong spin-flip scattering at the Py/Cu interface.
\end{abstract}


\maketitle



The performance of any device based on lateral spin valves depends on the magnitude of its signal.\cite{3,4} Similar to the effort put into every percentage increase for tunneling- or giant- magnetoresistance based devices\cite{4,5,6} the improvement of signal magnitude in lateral devices is an ongoing challenge. For any such attempt, one can control the materials used and the geometry of the device.\cite{6,7,8} Generally speaking, for lateral spin valves there are two important parameters. The first one is the polarization of injected current, which depends on the intrinsic spin polarization of the ferromagnetic (FM) material used\cite{3,4} and how effectively it can be injected into the non-magnetic electrode (NM)\cite{6,8,9,Valenz:apl}. The second parameter is the spin diffusion length of the NM.\cite{3,4,5,6,7,8} This is a measure of how far the injected spin imbalance diffuses inside the NM before it reaches an equilibrium state. In metallic lateral spin valves the devices are at the nanoscale, where the surface effects are comparable to the bulk. Thus the surface may have a significant influence on both the efficiency of injection and the spin diffusion length. 

In this letter, we show that the performance of Ni$_{80}$Fe$_{20}$ permalloy/copper (Py/Cu) lateral spin valves with transparent interfaces reduces considerably with decreasing the Cu thickness. We perform non-local spin-valve (NLSV) measurements\cite{3} for samples with different Cu thickness and multiple devices. This way, we are able to separately determine the spin diffusion length of the Cu electrode ($\lambda_{Cu}$) and the effective spin polarization of Py  ($\alpha_{Py}$).\cite{3} We find that both of these parameters increase with increasing Cu thickness. The thickness dependence of $\lambda_{Cu}$ and $\alpha_{Py}$ is explained by having different spin scattering mechanisms in the bulk of Cu, at the Cu surface and at the Py/Cu interface.

\begin{figure}
    \centering
       \includegraphics[width=0.37\textwidth]{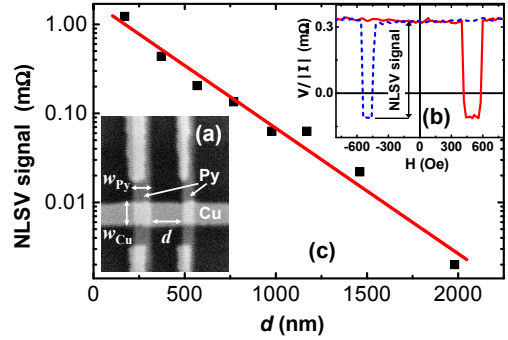}
    \caption{(color online) (a) Scanning electron micrograph of a typical device. Geometrical parameters are indicated in the image. (b) $V/|I|$ vs. magnetic field for the device with $d=370 nm$. Solid red (dotted blue) line is for increasing (decreasing) field. The NLSV signal is marked. (c) NLSV signal (black squares) vs. $d$ for the sample with $t_{Cu}=200 nm$. Red solid curve is the fit to Eq.~\ref{eq1}.}
    \label{fig1}
\end{figure}

The samples in this study were fabricated by a two-angle shadow evaporation technique. This enables the deposition of the Py and Cu electrodes on a Si substrate without breaking vacuum, which results in transparent and reproducible Py/Cu interfaces. The detailed fabrication process is described elsewhere.\cite{10} We prepared a series of samples with 6 to 8 lateral spin-valve devices on each sample (Fig.~\ref{fig1}(a)). Each device consists of a pair of Py electrodes crossed by a common Cu strip. Edge-to-edge distance ($d$) between pairs of Py electrodes is varied from 200 to 2000 nm. The Cu strip thickness ($t_{Cu}$) is varied between samples from 55 to 380 nm, and its width ($w_{Cu}$) is set to 250 nm for all samples. The thickness of the Py electrodes is fixed to 35 nm for all samples. In every device, the width of one Py electrode ($w_{Py}$)  is 100 nm and the other is 150 nm providing separate control over the magnetization of each electrode.\cite{11,13}

We measured our samples in a Helium-flow cryostat at 4.2 K. For each of the spin-valve devices we performed non-local measurements using a conventional DC reversal technique.\cite{14} The measured voltage depends on the relative orientation of magnetization of the two Py electrodes. It changes from a high value for parallel orientation of magnetization to a low value for antiparallel (see Fig.~\ref{fig1}(b)). The difference between the high and low voltages normalized to the current magnitude, $\Delta V/| I|$, is called the NLSV signal. The asymmetry between high and low values, which is theoretically not expected, is explained in Ref. \onlinecite{10}.

Figure \ref{fig1}(c) shows typical measurements of the NLSV signal (black squares) as a function of the edge-to-edge distance between Py electrodes for a sample with $t_{Cu}=200nm$. The NLSV signal decreases with increasing distance between electrodes. From this graph, we obtained the values of $\lambda_{Cu}$ and $\alpha_{Py}$, as described below. 

By applying the one-dimensional spin-diffusion model with transparent interfaces for our geometry,\cite{13,15,16} we write an expression for the NLSV signal as a function of the different geometrical and material parameters:
\begin{equation}
    \frac{\Delta V}{|I|}=\frac{2\alpha^2_{Py}R_{Cu}}{\left(2+\frac{R_{Cu}}{R_{Py}}\right)^2\exp\left(\frac{d}{\lambda_{Cu}}\right)-\left(\frac{R_{Cu}}{R_{Py}}\right)^2\exp\left(-\frac{d}{\lambda_{Cu}}\right)},
    \label{eq1}
\end{equation}
where $R_{Cu}=2\lambda_{Cu}/\sigma_{Cu}S_{Cu}$ and\\ $R_{Py}=2\lambda_{Py}/\sigma_{Py}S_{Py}\left(1-\alpha^2_{Py}\right)$  are spin-resistances of Cu and Py, respectively. $\lambda_{Py,Cu}$ are spin diffusion lengths, $\sigma_{Py,Cu}$ are the conductivities, and $S_{Py,Cu}$ are the cross-sectional areas of Py and Cu. For all samples, we use $\lambda_{Py}=5nm$\cite{3,6,17} and $\sigma_{Py} = 19 \mu\Omega cm$. $\sigma_{Py}$ was measured on a separate device deposited under nominally identical conditions, and is in agreement with values reported in the literature.\cite{3,7,17} All other variables, $\sigma_{Cu}$, $S_{Py}$, $S_{Cu}$ and $d$, were measured explicitly for each device. By fitting the data of each sample to Eq.~\ref{eq1} we extract the values of $\alpha_{Py}$ and $\lambda_{Cu}$ as the two fitting parameters. The resulting curve for the sample shown in Fig.~\ref{fig1}(c) is plotted as a red solid line, providing $\lambda_{Cu}=300\pm23 nm$ and $\alpha_{Py}=0.35\pm0.04$. These values are in agreement with other NLSV measurements of the Py/Cu system,\cite{3,7} but the spin polarization is lower than values obtained by other methods.\cite{5,6,GMR1,GMR2,PCAR3,PCAR2}

\begin{figure}
    \centering
       \includegraphics[width=0.37\textwidth]{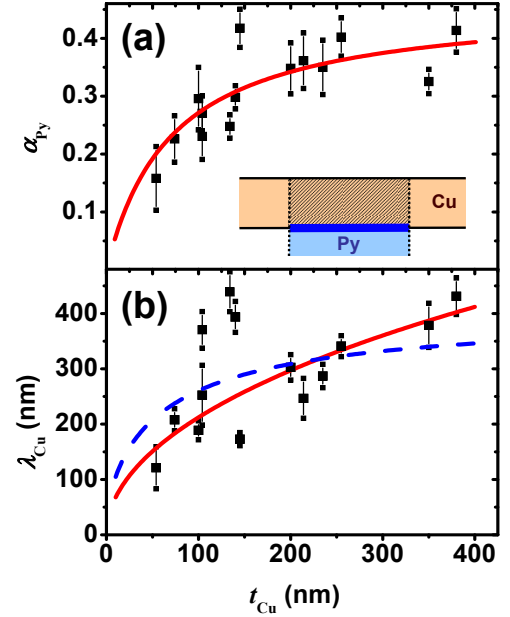}
    \caption{(color online) (a) $\alpha_{Py}$ (black squares) vs. $t_{Cu}$. Red solid curve is a fit to the model for injected spins (Eq.~\ref{eq3}). The inset is a schematic drawing showing a side view of the cross-shaped contact between the Cu strip (top) and the Py electrode (bottom). The shaded rectangle is the region of Cu strip right above the Py/Cu interface, which is marked by the thick line (blue online). (b) $\lambda_{Cu}$ (black squares) vs. $t_{Cu}$. Blue dashed line is a fitting curve for the modified spin diffusion model (Eq.~\ref{eq2}) accounting for all the Cu surfaces and red solid line assumes that only the top surface of Cu is relevant.}
    \label{fig2}
\end{figure}

In Fig.~\ref{fig2}(a) we plot $\alpha_{Py}$ (black squares) as a function of $t_{Cu}$. The spin polarization of Py increases with Cu thickness and starts saturating above $\sim 200 nm$. Theoretically, $\alpha_{Py}$ should be an intrinsic property of Py.\cite{13,15} Therefore, Fig.~\ref{fig2}(a) indicates that we are measuring an effective polarization of Py which is affected by the Cu thickness. Figure \ref{fig2}(b) depicts $\lambda_{Cu}$ (black squares) as a function of $t_{Cu}$. There is a general increase of the spin diffusion length of Cu for thicker samples, although there is some dispersion in the results. One should remember that each data point in this graph is obtained through a fitting of a different sample. Each sample was deposited separately and the exact geometry and deposition conditions could be the reason for this dispersion. The spin polarization plot has less dispersion implying that from deposition to deposition, the interface quality between Py and Cu is reproduced. 

\begin{figure}
    \centering
       \includegraphics[width=0.48\textwidth]{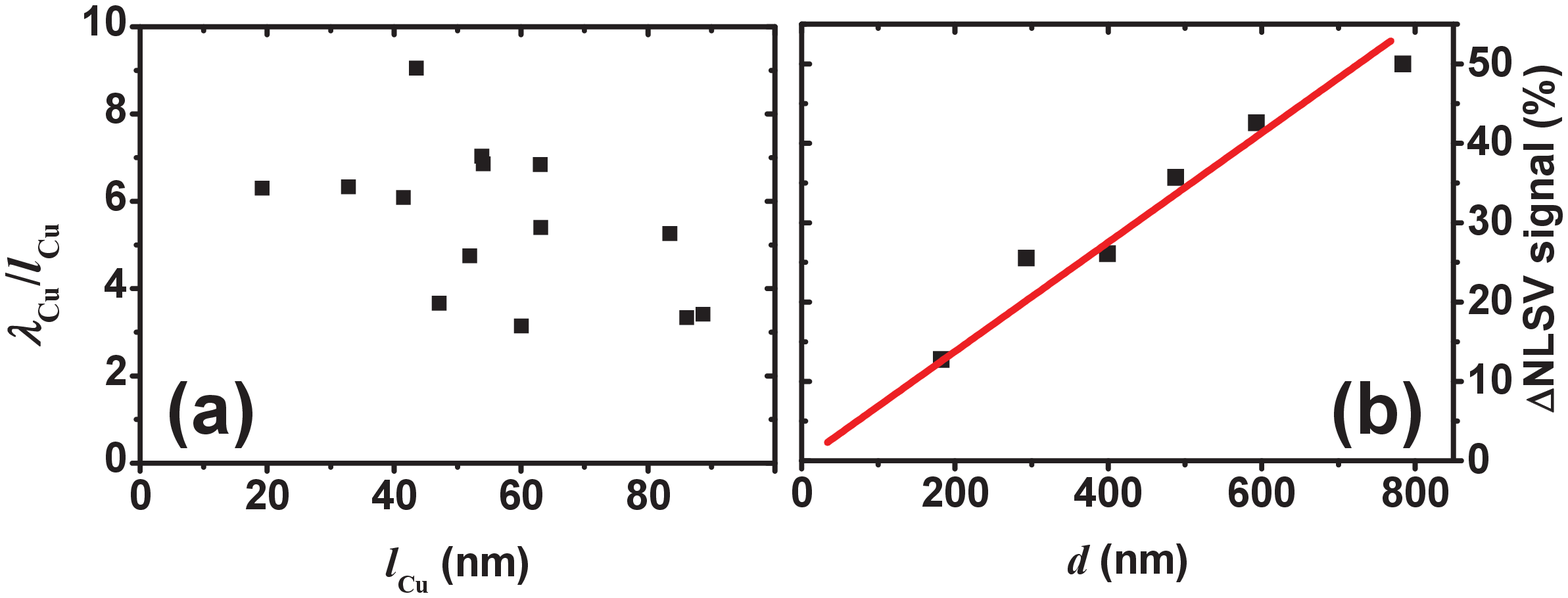}
    \caption{(color online) (a) $\lambda_{Cu}/\ell_{Cu}$ as a function of $\ell_{Cu}$. (b) Increase of NLSV signal after oxidation (black squares) vs. $d$ for a sample with $t_{Cu}=210 nm$. Red solid line demonstrates a linear behavior of the NLSV signal increase vs. $d$, indicating an increase in $\lambda_{Cu}$ after oxidation.}
    \label{fig3}
\end{figure}

To a first approximation, in the spin diffusion model, the probability of losing spin polarization of electrons is proportional to the time electrons spend diffusing inside the NM.\cite{6,13} Therefore, the only parameter that the spin diffusion length should be proportional to is the mean free path of electrons in the Cu strip ($\ell_{Cu}$),\cite{8} meaning that their ratio should be constant. We calculated $\ell_{Cu}$ from the resistivity of Cu measured for each sample.\cite{19,20} Fig.~\ref{fig3}(a) shows $\lambda_{Cu}/\ell_{Cu}$ as a function of $\ell_{Cu}$. We find that $\lambda_{Cu}/\ell_{Cu}$ is not constant, but fluctuates with $\ell_{Cu}$, implying that there is an additional spin-flip mechanism besides the one provided by momentum scattering. If this additional spin-flip scattering occurs at the surface then the spin diffusion length of Cu will change with thickness. We follow the derivation of the one-dimensional diffusion equation\cite{who:knows} to include surface effects. Spin flip scattering in the bulk is accounted for by assuming that for each momentum scattering there is a probability $p_b$ to also flip the spin. This model is modified by including a different probability $p_s$ to spin-flip at the surface. From here we can derive the dependence of $\lambda_{Cu}$ on the geometry of the Cu strip
\begin{equation}
    \lambda_{Cu}=\frac{\ell_{Cu}}{\sqrt{6p_b+\ell_{Cu}p_s\left(\frac{2}{t_{Cu}}+\frac{2}{w_{Cu}}\right)}}.
    \label{eq2}
\end{equation}
Using this formula, we fit our data of $\lambda_{Cu}$ vs. $t_{Cu}$ (blue dashed curve in Fig.~\ref{fig2}(b)), finding the scattering probabilities: $p_b=1.0\times10^{-4}\pm1.4\times10^{-3}$ and $p_s=0.036\pm0.024$. This fit follows the general trend of the experimental data. However, we obtain a better fit when we exclude spin-flip scattering from the side surfaces, indicating there is spin-scattering mainly from the top surface or the Cu/Si substrate interface (red solid curve in Fig.~\ref{fig2}(b)). Here the corresponding probabilities are $p_b=1.0\times10^{-4}\pm7.4\times10^{-4}$ and $p_s=0.14\pm0.06$. In both cases the results show that the probability to scatter spins from the surface is $\sim3$ orders of magnitude higher than the bulk. Therefore, the spin diffusion length of Cu is dominated by surface scattering. Enhanced spin-flip scattering from metal surfaces was suggested earlier by other groups \cite{17,vran,pierre:prl} and it was also found in recent theoretical calculations.\cite{21}

We were able to attain additional information by oxidizing a few samples and re-measuring the NLSV signal for the devices. In Fig.~\ref{fig3}(b) we show the percentage difference between NLSV signals measured before and after oxidation for the $210nm$-thick sample. The NLSV signal increases for each device, and the extracted $\lambda_{Cu}$ increases as well from $247\pm36 nm$ to $282\pm48 nm$. Since the Cu/Si interface is not affected by the oxidation, the increase in $\lambda_{Cu}$ indicates that it is the top surface which provides strong spin-flip scattering.

We apply the modified spin-diffusion model to understand the decrease of $\alpha_{Py}$ for thinner Cu. The dependence of $\alpha_{Py}$ on $t_{Cu}$ is not explained by the one-dimensional spin-diffusion model because in this model the spin polarization is an intrinsic parameter of the FM.

In the actual sample, there is a region of the Cu strip above the Py/Cu interface (see inset of Fig.~\ref{fig2}(a)) that is not taken into account by the one-dimensional model. This region also provides spin-flip scattering for the injected electrons, where one of the scattering surfaces is a FM. As a result, the injected electrons lose some of their spin polarization before continuing to diffuse along the Cu strip. Therefore, $\alpha_{Py}$ is an effective spin polarization and it is smaller than the intrinsic one.

We calculate the effective polarization of the injected electrons by considering the spin-scattering contributions from the bulk of Cu, the Cu surface and the Cu/Py interface. If $N_{In}$ is the number of injected polarized electron per unit time (in our model there is no spin scattering occurs when those electrons cross the interface), and $N_{Out}$ is the number leaving the area above the injector then $\alpha_{Py}=\alpha_{int}\left(N_{Out}/N_{In}\right)$ where $\alpha_{int}$ is the intrinsic polarization of Py. Now, $N_{Out}$ equals $N_{In}$ minus the number of spins per unit time which scatter above the Py electrode.  These include spins scattered at the Cu surface, in the bulk and at the interface, and are functions of the scattering probabilities $p_s$, $p_b$ and $p_i$ (the spin-flip probability for momentum scattering at the Py/Cu interface), the geometrical parameters ($t_{Cu}$, $w_{Cu}$, $w_{Py}$) and $\ell_{Cu}$. For a steady state we attain:
\begin{equation} \alpha_{Py}=\frac{\alpha_{int}}{1+p_b\frac{3w_{Py}}{\ell_{Cu}}+\frac{w_{Py}}{2}\left(\frac{p_s+p_i}{t_{Cu}}+\frac{2p_s}{w_{Cu}}\right)}.
    \label{eq3}
\end{equation}
We fit the spin polarization data to Eq.~\ref{eq3} using previously obtained $p_s$ and $p_b$, and $w_{Py}$ measured separately for each device.
We attained an excellent fit (red solid curve in Fig.~\ref{fig2}(a)), with $p_i=1.0\pm0.4$ and $\alpha_{int}=0.50\pm0.07$. That the fit quality and parameters do not depend on whether we include all free Cu surfaces or only the top surface. The value $p_i=1$ is reasonable, since it means that every time an electron scatters at the magnetic Py/Cu interface it completely loses its spin information. More important, the value $\alpha_{int}=0.50$ is in agreement with values obtained by other methods (0.35-0.8).\cite{PCAR3,PCAR2,GMR1,GMR2}
  
To summarize, we found that both the spin diffusion length of Cu and the effective spin polarization of Py are strongly dependent on the Cu thickness in Py/Cu lateral spin valves devices with transparent contacts. The spin diffusion length of Cu is reduced for thin Cu strips due to dominant spin-flip scattering at the surface. In addition, we find that the effective spin polarization is strongly affected by the presence of the Py/Cu interface when the Cu thickness is less than 200 nm. The origin of the spin-flip scattering at the surface is still not understood and needs more attention.\cite{21} Nevertheless, we show that increasing the thickness of the Cu strip is an effective way of improving the performance of lateral spin valve devices.

This work was supported and funded by the US DOE.


\begin{thebibliography}{25}
\expandafter\ifx\csname natexlab\endcsname\relax\def\natexlab#1{#1}\fi
\expandafter\ifx\csname bibnamefont\endcsname\relax
  \def\bibnamefont#1{#1}\fi
\expandafter\ifx\csname bibfnamefont\endcsname\relax
  \def\bibfnamefont#1{#1}\fi
\expandafter\ifx\csname citenamefont\endcsname\relax
  \def\citenamefont#1{#1}\fi
\expandafter\ifx\csname url\endcsname\relax
  \def\url#1{\texttt{#1}}\fi
\expandafter\ifx\csname urlprefix\endcsname\relax\def\urlprefix{URL }\fi
\providecommand{\bibinfo}[2]{#2}
\providecommand{\eprint}[2][]{\url{#2}}

\bibitem[{\citenamefont{Jedema et~al.}(2001)\citenamefont{Jedema, Filip, and
  van Wees}}]{3}
\bibinfo{author}{\bibfnamefont{F.~J.} \bibnamefont{Jedema}},
  \bibinfo{author}{\bibfnamefont{A.~T.} \bibnamefont{Filip}}, \bibnamefont{and}
  \bibinfo{author}{\bibfnamefont{B.~J.} \bibnamefont{van Wees}},
  \bibinfo{journal}{Nature} \textbf{\bibinfo{volume}{410}},
  \bibinfo{pages}{345} (\bibinfo{year}{2001}).

\bibitem[{\citenamefont{Zutic et~al.}(2004)\citenamefont{Zutic, Fabian, and
  Sarma}}]{4}
\bibinfo{author}{\bibfnamefont{I.}~\bibnamefont{Zutic}},
  \bibinfo{author}{\bibfnamefont{J.}~\bibnamefont{Fabian}}, \bibnamefont{and}
  \bibinfo{author}{\bibfnamefont{S.~D.} \bibnamefont{Sarma}},
  \bibinfo{journal}{Rev. Mod. Phys.} \textbf{\bibinfo{volume}{76}},
  \bibinfo{pages}{323} (\bibinfo{year}{2004}).

\bibitem[{\citenamefont{Gijs and Bauer}(1997)}]{5}
\bibinfo{author}{\bibfnamefont{M.~A.~M.} \bibnamefont{Gijs}} \bibnamefont{and}
  \bibinfo{author}{\bibfnamefont{G.~E.~W.} \bibnamefont{Bauer}},
  \bibinfo{journal}{Advances in Physics} \textbf{\bibinfo{volume}{46}},
  \bibinfo{pages}{285 } (\bibinfo{year}{1997}).

\bibitem[{\citenamefont{Bass and Pratt}(2007)}]{6}
\bibinfo{author}{\bibfnamefont{J.}~\bibnamefont{Bass}} \bibnamefont{and}
  \bibinfo{author}{\bibfnamefont{W.~P.} \bibnamefont{Pratt}},
  \bibinfo{journal}{J. Phys.: Condens. Matter} \textbf{\bibinfo{volume}{19}},
  \bibinfo{pages}{183201} (\bibinfo{year}{2007}).

\bibitem[{\citenamefont{Kimura and Otani}(2007)}]{7}
\bibinfo{author}{\bibfnamefont{T.}~\bibnamefont{Kimura}} \bibnamefont{and}
  \bibinfo{author}{\bibfnamefont{Y.}~\bibnamefont{Otani}}, \bibinfo{journal}{J.
  Phys.: Condens. Matter} \textbf{\bibinfo{volume}{19}},
  \bibinfo{pages}{165216} (\bibinfo{year}{2007}).

\bibitem[{\citenamefont{Poli et~al.}(2006)\citenamefont{Poli, Urech,
  Korenivski, and Haviland}}]{8}
\bibinfo{author}{\bibfnamefont{N.}~\bibnamefont{Poli}},
  \bibinfo{author}{\bibfnamefont{M.}~\bibnamefont{Urech}},
  \bibinfo{author}{\bibfnamefont{V.}~\bibnamefont{Korenivski}},
  \bibnamefont{and} \bibinfo{author}{\bibfnamefont{D.~B.}
  \bibnamefont{Haviland}}, \bibinfo{journal}{J. Appl. Phys.}
  \textbf{\bibinfo{volume}{99}}, \bibinfo{pages}{136601}
  (\bibinfo{year}{2006}).

\bibitem[{\citenamefont{Fert and Lee}(1996)}]{9}
\bibinfo{author}{\bibfnamefont{A.}~\bibnamefont{Fert}} \bibnamefont{and}
  \bibinfo{author}{\bibfnamefont{S.-F.} \bibnamefont{Lee}},
  \bibinfo{journal}{Phys. Rev. B} \textbf{\bibinfo{volume}{53}},
  \bibinfo{pages}{6554} (\bibinfo{year}{1996}).

\bibitem[{\citenamefont{Valenzuela and Tinkham}(2004)}]{Valenz:apl}
\bibinfo{author}{\bibfnamefont{S.~O.} \bibnamefont{Valenzuela}}
  \bibnamefont{and} \bibinfo{author}{\bibfnamefont{M.}~\bibnamefont{Tinkham}},
  \bibinfo{journal}{Appl. Phys. Lett.} \textbf{\bibinfo{volume}{85}},
  \bibinfo{pages}{5914} (\bibinfo{year}{2004}).

\bibitem[{\citenamefont{Casanova et~al.}(2009)\citenamefont{Casanova, Sharoni,
  Erekhinsky, and Schuller}}]{10}
\bibinfo{author}{\bibfnamefont{F.}~\bibnamefont{Casanova}},
  \bibinfo{author}{\bibfnamefont{A.}~\bibnamefont{Sharoni}},
  \bibinfo{author}{\bibfnamefont{M.}~\bibnamefont{Erekhinsky}},
  \bibnamefont{and} \bibinfo{author}{\bibfnamefont{I.~K.}
  \bibnamefont{Schuller}}, \bibinfo{journal}{Phys. Rev. B}
  \textbf{\bibinfo{volume}{79}}, \bibinfo{pages}{184415}
  (\bibinfo{year}{2009}).

\bibitem[{\citenamefont{Hong and Giordano}(1995)}]{11}
\bibinfo{author}{\bibfnamefont{K.}~\bibnamefont{Hong}} \bibnamefont{and}
  \bibinfo{author}{\bibfnamefont{N.}~\bibnamefont{Giordano}},
  \bibinfo{journal}{Phys. Rev. B} \textbf{\bibinfo{volume}{51}},
  \bibinfo{pages}{9855} (\bibinfo{year}{1995}).

\bibitem[{\citenamefont{Jedema et~al.}(2003)\citenamefont{Jedema, Nijboer,
  Filip, and van Wees}}]{13}
\bibinfo{author}{\bibfnamefont{F.~J.} \bibnamefont{Jedema}},
  \bibinfo{author}{\bibfnamefont{M.~S.} \bibnamefont{Nijboer}},
  \bibinfo{author}{\bibfnamefont{A.~T.} \bibnamefont{Filip}}, \bibnamefont{and}
  \bibinfo{author}{\bibfnamefont{B.~J.} \bibnamefont{van Wees}},
  \bibinfo{journal}{Phys. Rev. B} \textbf{\bibinfo{volume}{67}},
  \bibinfo{pages}{085319} (\bibinfo{year}{2003}).

\bibitem[{\citenamefont{Daire et~al.}(2005)\citenamefont{Daire, Goeke, and
  Tupta}}]{14}
\bibinfo{author}{\bibfnamefont{A.}~\bibnamefont{Daire}},
  \bibinfo{author}{\bibfnamefont{W.}~\bibnamefont{Goeke}}, \bibnamefont{and}
  \bibinfo{author}{\bibfnamefont{M.~A.} \bibnamefont{Tupta}},
  \emph{\bibinfo{title}{White paper: New instruments can lock out lock-ins;
  http://www.keithley.com/data?asset=50379}} (\bibinfo{year}{2005}).

\bibitem[{\citenamefont{Kimura et~al.}(2005)\citenamefont{Kimura, Hamrle, and
  Otani}}]{15}
\bibinfo{author}{\bibfnamefont{T.}~\bibnamefont{Kimura}},
  \bibinfo{author}{\bibfnamefont{J.}~\bibnamefont{Hamrle}}, \bibnamefont{and}
  \bibinfo{author}{\bibfnamefont{Y.}~\bibnamefont{Otani}},
  \bibinfo{journal}{Phys. Rev. B} \textbf{\bibinfo{volume}{72}},
  \bibinfo{pages}{014461} (\bibinfo{year}{2005}).

\bibitem[{\citenamefont{Valet and Fert}(1993)}]{16}
\bibinfo{author}{\bibfnamefont{T.}~\bibnamefont{Valet}} \bibnamefont{and}
  \bibinfo{author}{\bibfnamefont{A.}~\bibnamefont{Fert}},
  \bibinfo{journal}{Phys. Rev. B} \textbf{\bibinfo{volume}{48}},
  \bibinfo{pages}{7099} (\bibinfo{year}{1993}).

\bibitem[{\citenamefont{Kimura et~al.}(2008)\citenamefont{Kimura, Sato, and
  Otani}}]{17}
\bibinfo{author}{\bibfnamefont{T.}~\bibnamefont{Kimura}},
  \bibinfo{author}{\bibfnamefont{T.}~\bibnamefont{Sato}}, \bibnamefont{and}
  \bibinfo{author}{\bibfnamefont{Y.}~\bibnamefont{Otani}},
  \bibinfo{journal}{Phys. Rev. Lett.} \textbf{\bibinfo{volume}{100}},
  \bibinfo{pages}{066602} (\bibinfo{year}{2008}).

\bibitem[{\citenamefont{Dubois et~al.}(1999)\citenamefont{Dubois, Piraux,
  George, Ounadjela, Duvail, and Fert}}]{GMR1}
\bibinfo{author}{\bibfnamefont{S.}~\bibnamefont{Dubois}},
  \bibinfo{author}{\bibfnamefont{L.}~\bibnamefont{Piraux}},
  \bibinfo{author}{\bibfnamefont{J.~M.} \bibnamefont{George}},
  \bibinfo{author}{\bibfnamefont{K.}~\bibnamefont{Ounadjela}},
  \bibinfo{author}{\bibfnamefont{J.~L.} \bibnamefont{Duvail}},
  \bibnamefont{and} \bibinfo{author}{\bibfnamefont{A.}~\bibnamefont{Fert}},
  \bibinfo{journal}{Phys. Rev. B} \textbf{\bibinfo{volume}{60}},
  \bibinfo{pages}{477} (\bibinfo{year}{1999}).

\bibitem[{\citenamefont{Holody et~al.}(1998)\citenamefont{Holody, Chiang,
  Loloee, Bass, Pratt, and Schroeder}}]{GMR2}
\bibinfo{author}{\bibfnamefont{P.}~\bibnamefont{Holody}},
  \bibinfo{author}{\bibfnamefont{W.~C.} \bibnamefont{Chiang}},
  \bibinfo{author}{\bibfnamefont{R.}~\bibnamefont{Loloee}},
  \bibinfo{author}{\bibfnamefont{J.}~\bibnamefont{Bass}},
  \bibinfo{author}{\bibfnamefont{W.~P.} \bibnamefont{Pratt}}, \bibnamefont{and}
  \bibinfo{author}{\bibfnamefont{P.~A.} \bibnamefont{Schroeder}},
  \bibinfo{journal}{Phys. Rev. B} \textbf{\bibinfo{volume}{58}},
  \bibinfo{pages}{12230} (\bibinfo{year}{1998}).

\bibitem[{\citenamefont{Soulen Jr~\textit{et al.}}(1998)}]{PCAR3}
\bibinfo{author}{\bibfnamefont{R.~J.} \bibnamefont{Soulen Jr~\textit{et al.}}},
  \bibinfo{journal}{Science} \textbf{\bibinfo{volume}{282}},
  \bibinfo{pages}{85} (\bibinfo{year}{1998}).

\bibitem[{\citenamefont{Nadgorny~\textit{et al.}}(2000)}]{PCAR2}
\bibinfo{author}{\bibfnamefont{B.}~\bibnamefont{Nadgorny~\textit{et al.}}},
  \bibinfo{journal}{Phys. Rev. B} \textbf{\bibinfo{volume}{61}},
  \bibinfo{pages}{R3788} (\bibinfo{year}{2000}).

\bibitem[{\citenamefont{Jedema}(2002)}]{19}
\bibinfo{author}{\bibfnamefont{F.~J.} \bibnamefont{Jedema}}, Ph.D. thesis,
  \bibinfo{school}{Rijksuniversiteit Groningen} (\bibinfo{year}{2002}).

\bibitem[{\citenamefont{Abrikosov}(1988)}]{20}
\bibinfo{author}{\bibfnamefont{A.}~\bibnamefont{Abrikosov}},
  \emph{\bibinfo{title}{Fundamentals of the Theory of Metals}}
  (\bibinfo{publisher}{Elsevier Science Publishers B.V.},
  \bibinfo{address}{Amsterdam}, \bibinfo{year}{1988}).

\bibitem[{\citenamefont{Glicksman}(2000)}]{who:knows}
\bibinfo{author}{\bibfnamefont{M.~E.} \bibnamefont{Glicksman}},
  \emph{\bibinfo{title}{Diffusion in Solids}} (\bibinfo{publisher}{Wiley},
  \bibinfo{address}{New York}, \bibinfo{year}{2000}).

\bibitem[{\citenamefont{Vranken et~al.}(1988)\citenamefont{Vranken,
  Van~Haesendonck, and Bruynseraede}}]{vran}
\bibinfo{author}{\bibfnamefont{J.}~\bibnamefont{Vranken}},
  \bibinfo{author}{\bibfnamefont{C.}~\bibnamefont{Van~Haesendonck}},
  \bibnamefont{and}
  \bibinfo{author}{\bibfnamefont{Y.}~\bibnamefont{Bruynseraede}},
  \bibinfo{journal}{Physical Review B} \textbf{\bibinfo{volume}{37}},
  \bibinfo{pages}{8502} (\bibinfo{year}{1988}).

\bibitem[{\citenamefont{Pierre and Birge}(2002)}]{pierre:prl}
\bibinfo{author}{\bibfnamefont{F.}~\bibnamefont{Pierre}} \bibnamefont{and}
  \bibinfo{author}{\bibfnamefont{N.~O.} \bibnamefont{Birge}},
  \bibinfo{journal}{Physical Review Letters} \textbf{\bibinfo{volume}{89}},
  \bibinfo{pages}{206804} (\bibinfo{year}{2002}).

\bibitem[{\citenamefont{Suhl}(work in progress)}]{21}
\bibinfo{author}{\bibfnamefont{H.}~\bibnamefont{Suhl}},
  \emph{\bibinfo{title}{Enhanced spin-flip scattering at the surface of
  metals}} (\bibinfo{year}{work in progress}).

\end{thebibliography}

\end{document}